\documentstyle[sprocl,epsf]{article}

\def\be{\begin{equation}}
\def\ee{\end{equation}}
\def\bea{\begin{eqnarray}}
\def\eea{\end{eqnarray}}

\def\gsim{\mathrel{\vcenter{\hbox{$>$}\nointerlineskip\hbox{$\sim$}}}}

\begin{document}

\title{THE LIGHTEST NEUTRAL AND DOUBLY CHARGED HIGGS BOSONS OF
SUPERSYMMETRIC LEFT-RIGHT MODELS \footnote{Talk presented by
K. Huitu in the International Workshop on Linear Colliders,
Sitges, April 28 - May 5, 1999.}}

\author{ K. HUITU$^1$, P.N. PANDITA$^{1,2}$, K. PUOLAM\"AKI$^1$ }

\address{$^1$Helsinki Institute of Physics, P.O.B. 9,\\
FIN-00014 University of Helsinki, Finland}
\address{$^2$Department of Physics, North Eastern Hill University,\\
Shillong 793022, India}

\maketitle\abstracts{
We review the phenomenology of  light Higgs scalars in 
supersymmetric left-right models.
We consider models with minimal 
particle content (with and without non-renormalizable 
higher-dimensional terms)
and with additional Higgs superfields.
The upper bound on the lightest $CP$-even neutral 
Higgs boson in these models is larger than
in the minimal supersymmetric standard model,
and the Higgs couplings to fermions approach those of the Standard Model.
Possibly light doubly charged Higgs boson
may provide the best signature of these models.}

\section{The models\label{sec:models}}

The left-right models are interesting  for many
reasons, e.g. they provide a natural way to generate light masses for
the neutrinos via the see-saw mechanism
\cite{Gell-Mann:1979}.
An important motivation for the
supersymmetric left-right models $^{2-14}$ is due to the fact
\cite{Mohapatra:1986su,Martin:1992mq} that if the gauge symmetry
is extended to $SU(2)_L \times U(1)_{I_{3R}} \times
U(1)_{B-L}$, or to $SU(2)_L \times SU(2)_R \times U(1)_{B-L}$, then 
R-parity is conserved in the Lagrangian of the theory.
Thus one of the major problematic features of the MSSM is resolved by
a gauge symmetry.
Here we will concentrate on the model with the $SU(2)_R$ symmetry,
the supersymmetric left-right model (SLRM).

While the problem of R-parity is solved, the particle content of the model
is enlarged.
In addition to the new superfields containing the gauge bosons 
of the $SU(2)_R$ symmetry, one has a right-handed neutrino
superfield ($\nu_L^c$).
The Higgs sector in the SLRM is chosen to have
triplets in the spectrum, in which case one can have the conventional 
see-saw mechanism for neutrino mass generation.
The $SU(2)_L$ will be broken mainly by bidoublets which contain the
doublets of the MSSM.
Thus, the Higgs sector consists of the 
following superfields:
\begin{eqnarray}
\Phi = \left( \begin{array}{cc} \Phi^0_1 & \Phi^+_1 \\ \Phi^-_2 &
\Phi^0_2 \end{array} \right),\;  \chi = \left(
\begin{array}{cc} \chi^0_1 & \chi^+_1 \\ \chi^-_2 & \chi^0_2
\end{array} \right) \sim (1,2,2,0) ,&& \nonumber \\ 
\Delta_R  
\sim (1,1,3,-2),\;  \delta_R 
\sim (1,1,3,2), \; \Delta_L 
\sim (1,3,1,-2),\;  \delta_L   \sim
(1,3,1,2) .&&
\label{eq:fields2}
\end{eqnarray}
The $SU(2)_L$ triplets $\Delta_L$ and $\delta_L$  
make the Lagrangian fully symmetric under $L \leftrightarrow
R$.
Left triplets are not needed
for symmetry breaking or the see-saw mechanism.

The VEVs preserving the $U(1)_{\rm{em}}$ 
gauge invariance can be written as
\begin{eqnarray}
&&\langle \Phi \rangle = \left( \begin{array}{cc} \kappa_1 & 0 \\ 0 &
e^{i \phi_1}\kappa_1' \end{array} \right) ,\;  
\langle \chi \rangle =\left(\begin{array}{cc} e^{i \phi_2} \kappa_2' &
0 \\ 0 & \kappa_2\end{array} \right) , \;
\langle \tilde\nu_L \rangle = \sigma_L ,\;  
\langle\tilde\nu_L^c\rangle =  \sigma_R,\nonumber \\ 
&& 
\langle \Delta_R^0 \rangle = v_{\Delta_R} ,\;  
\langle \delta_R^0 \rangle = v_{\delta_R} ,\;
 \langle \Delta_L^0 \rangle =v_{\Delta_L}  ,\;  
\langle \delta_L^0 \rangle = v_{\delta_L} , \;
\label{eq:vevs}
\end{eqnarray}
The triplet VEVs $v_{\Delta_R,\delta_R}$ are,
according to the lower bounds \cite{Caso} on heavy
W- and Z-boson masses, in the range $v_{\Delta_R,\delta_R} \gsim 1$
TeV.  
The VEVs $\kappa_{1,2}'$ contribute to the mixing of the charged gauge bosons
and to the flavour changing neutral currents, and are usually assumed
to vanish. 
The left-triplet VEVs $v_{\Delta_L,\delta_L}$
must be small,  since the electroweak $\rho$
parameter is
close to unity, $\rho = 0.9998 \pm 0.0008$ \cite{Caso}.
With the minimal field content and renormalizable model, 
the only way to preserve the
$U(1)_{em}$ gauge symmetry is to break the R-parity by a sneutrino VEV
\cite{Kuchimanchi:1993jg,Huitu:1995zm}.

An alternative to the minimal left-right supersymmetric
model 
involves  additional  triplet fields, 
$\Omega_L(1,3,1,0)$ and $\Omega_R(1,1,3,0)$  \cite{Aulakh:1997ba}.
In these extended models 
the gauge group $SU(2)_R \times U(1)_{B-L}$ is
broken first to an intermediate symmetry group
$U(1)_R \times U(1)_{B-L}$, and at the
second stage   to $U(1)_Y$ at
a lower scale. In this theory the parity-breaking minimum
respects the electromagnetic gauge invariance without a sneutrino VEV.

A second option is to add non-renormalizable terms
to the Lagrangian of the minimal model
\cite{Martin:1992mq,AMRS,AMS}.
It has been shown that the addition of 
terms suppressed by a high scale such as Planck mass,
$M\sim 10^{19}$ GeV, with the minimal field content
ensures the correct pattern of symmetry breaking
in the SLRM with the intermediate
scale $M_R \gsim 10^{10} - 10^{11}$ GeV, and $R$-parity remains
exact.

\section{The upper limit on the lightest CP-even Higgs\label{sec:h0mass}}

In the case of the SLRM we have many
new couplings and also new scales in the model and it is
not obvious, what is the  upper limit on the lightest CP-even Higgs 
boson mass.
This mass bound is a very important issue, 
since the experiments are  approaching
the upper limit of the lightest Higgs boson mass in the MSSM.

A general method to find an upper limit for the lightest Higgs mass  
was presented in \cite{Comelli:1996xg}.
This method has been applied to the mass of the lightest Higgs, 
$m_{h}$, of SLRM \cite{HPP3} in three cases:
(A) $R$-parity is spontaneously broken (sneutrinos get VEVs),
(B) $R$-parity is conserved because of additional triplets, and
(C) $R$-parity is conserved because of nonrenormalizable terms.

For the minimal model, case (A), the upper bound on $m_{h}$ is  
\cite{HPP3}
\bea
m_h^2 \leq \frac 1{2 v^2} \left[ g_L^2 (\omega_\kappa^2+\sigma_L^2)^2+
g_R^2 \omega_\kappa^4+g_{B-L}^2 \sigma_L^4 
+ 8 (h_{\Phi L}
\kappa_1'+h_{\chi L} \kappa_2 )^2 \sigma_L^2 + 8 h_{\Delta_L}^2
\sigma_L^4 \right] ,
\label{eq:treeupper2}
\eea
where
$v^2 = \kappa_1^2+ \kappa_1'^2+\kappa_2^2+\kappa_2'^2+\sigma_L^2$ and
$\omega_\kappa^2 = \kappa_1^2-\kappa_2^2-\kappa_1'^2+\kappa_2'^2 .$
The addition of extra triplets does not change 
this bound.
Thus, the bound for the case (B), can be obtained from
(\ref{eq:treeupper2}) by taking the limit $\sigma_L\rightarrow 0$.
The total number of nonrenormalizable terms in case (C) is
large.
However, the contribution to the Higgs mass bound from these
terms is found to be \cite{HPP3}
typically numerically negligible.
Therefore the upper bound for this class of models is essentially the
same as in the case (B).

The radiative corrections to the lightest Higgs
mass are significant.
For the SLRM lightest Higgs they have been calculated in detail
\cite{HPP3}.
For nearly degenerate stop masses, 
the radiative corrections on $m_h$  in the SLRM 
differ in form from the MSSM upper bound only because of
new supersymmetric Higgs mixing parameters. 
\begin{figure}[t]
\leavevmode
\begin{center}
\mbox{\epsfxsize=5.truecm\epsfysize=5.truecm\epsffile{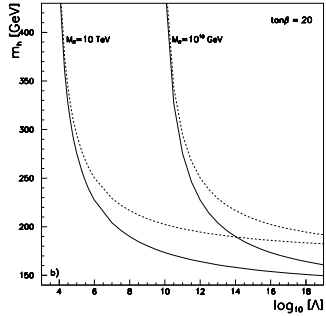}}
\end{center}
\caption{\label{mh1}The upper bound on the mass 
of the lightest neutral Higgs boson.
The bi- and trilinear soft supersymmetry breaking parameters are  1 TeV
(solid line) and 10 TeV (dashed line).}
\end{figure}

The upper bound on 
the mass of the lightest Higgs is plotted in Fig.\ref{mh1} 
as a function of the scale $\Lambda $ up to
which the SLRM remains
perturbative.  
The upper bound is shown for two different values of
the $SU(2)_R$ breaking
scale, $M_R=10$ TeV and $M_R=10^{10}$ GeV,
and for two values of soft supersymmetry breaking
mass parameter, $M_s=1$ TeV and $M_s=10$ TeV.
For large values of $\Lambda $ the upper bound 
is below 200 GeV.

\subsection{Couplings of the lightest neutral Higgs to fermions 
in the SLRM}

In order to study the phenomenology of the lightest Higgs boson
in the SLRM, its couplings to fermions are needed.

In the left-right symmetric models problems with
FCNC are expected if several light Higgs bosons exist \cite{EGN}
unless  $m_{H_{FCNC}}\gsim {\cal{O}}(1$ TeV). 
Thus the relevant limit to discuss is the one in which all the neutral Higgs
bosons, except the lightest one, are heavy.
It has been shown that in the decoupling limit the 
Yukawa couplings of the $\tau$'s are the same in the SM and the SLRM 
even if the $\tau$'s contain a large fraction of gauginos or
higgsinos \cite{HPP3}.

\section{The lightest doubly charged Higgs\label{sec:hpp}}

In addition to the lightest neutral CP-even Higgs, it has been known
for quite some time \cite{Huitu:1995zm} that the lightest doubly 
charged Higgs boson in these models may be light.
Whether it is observable in the experiments is an interesting 
issue, since
this particle may both reveal the nature
of the gauge group and help to determine the particular supersymmetric
left-right model in question.

There are four doubly charged Higgs bosons in the SLRM, of which
two are right-handed and two left-handed.
The masses of the left-handed triplets are expected to be of the same 
order as the soft terms.
The mass matrix for the right-handed triplets depends on the
right-triplet VEV.
Nevertheless,
it was noticed in \cite{Huitu:1995zm} that in the SLRM with broken R-parity
one right-handed doubly charged scalar tends to be light.
Also,  in the nonrenormalizable case
it is possible to have light doubly charged scalars \cite{CM58}.
On the other hand, in the nonsupersymmetric left-right model all the
doubly charged scalars typically have a mass of the order of the 
right-handed scale \cite{GGMKO}.
This is also true in the SLRM with enlarged particle content
\cite{AMRS}.
Thus a light doubly charged Higgs would be a strong indication of a
supersymmetric left-right model with minimal particle content.

\begin{figure}[t]
\leavevmode
\begin{center}
\mbox{\epsfxsize=5.truecm\epsfysize=5.truecm\epsffile{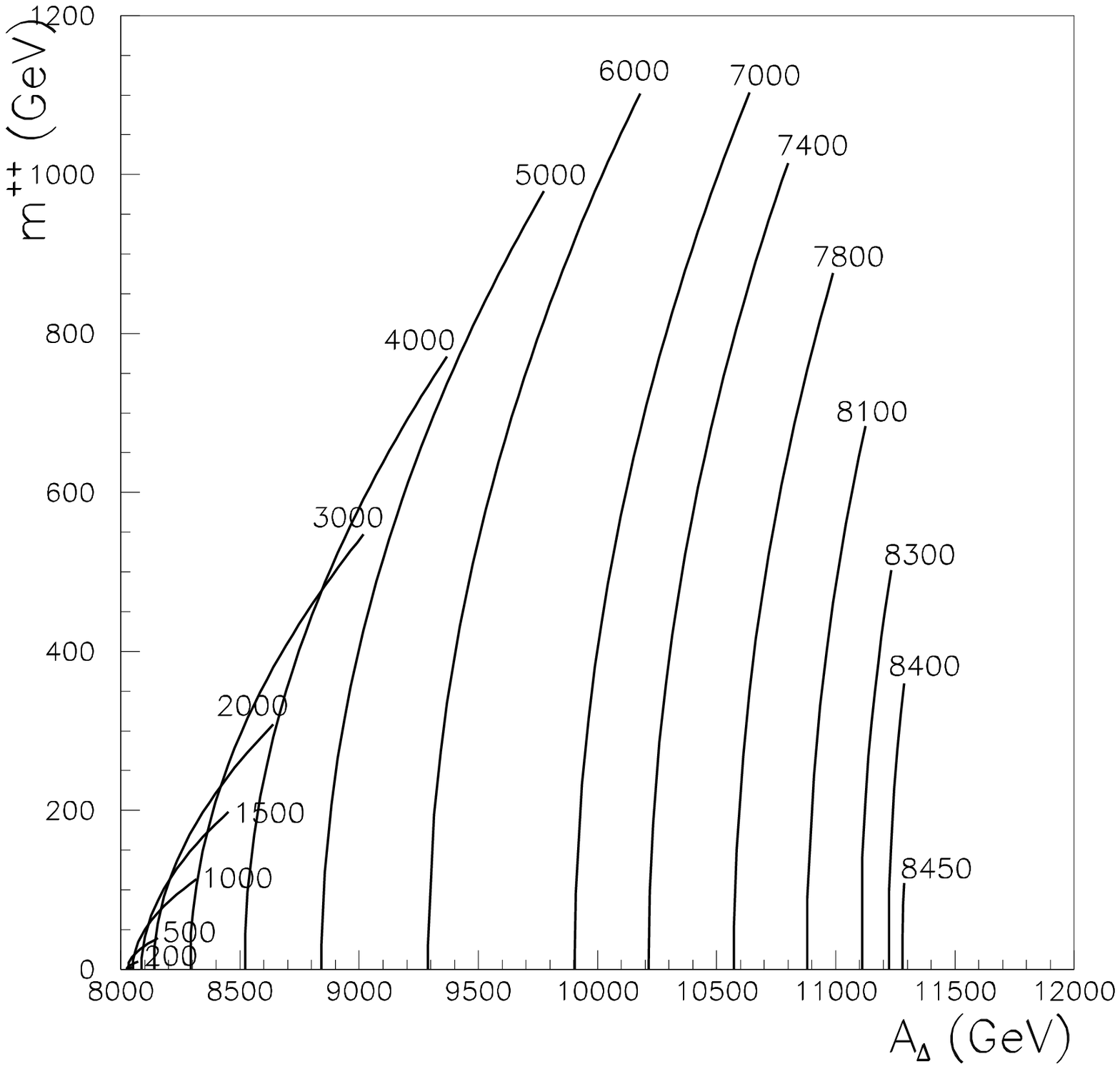}
\epsfxsize=5.truecm\epsfysize=5.truecm\epsffile{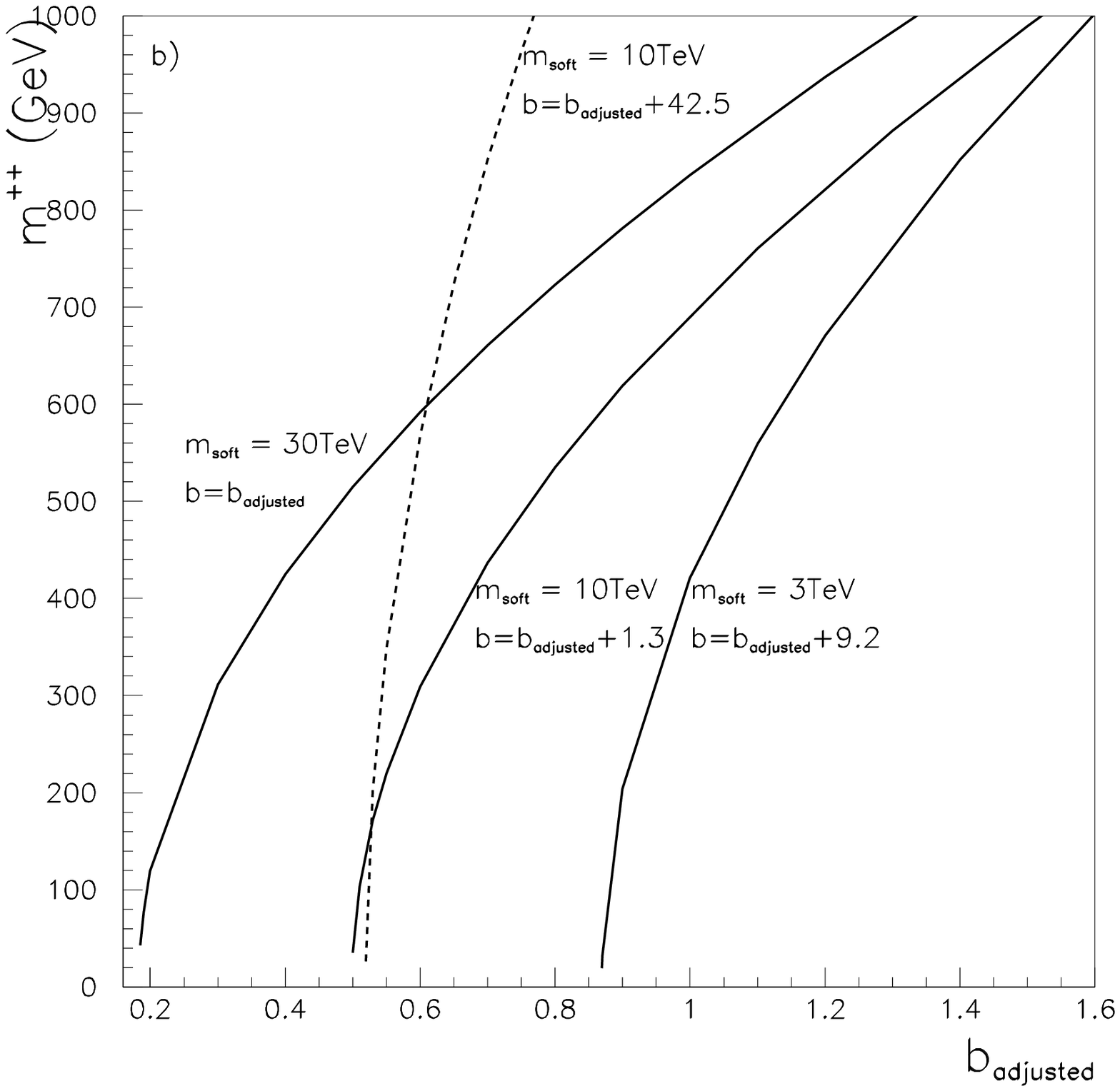}}
\end{center}
\caption{\label{mpp_ad}
The mass $m_{H^{++}}$ of the lightest doubly charged Higgs.
In a) the mass is as a function of the soft trilinear coupling $A_\Delta$.
$\sigma_R$ varies in the allowed range of 100 GeV to 8.45 TeV.
In b) the mass is as a function of the
nonrenormalizable $b_R$-parameter.
In b)
$v_R^2/M=10^{2}$ GeV and 
$D =(3$ TeV)$^2$ (solid line).  For 
$m_{soft} =10$ TeV also $D=10$ TeV$^2$ is shown (dashed line).
The soft supersymmetry breaking parameters and the $b_R$ parameters
are marked in the figure.
$\tan\beta=50$.}
\end{figure}

In Figure \ref{mpp_ad} a) an example of $H^{++}$ masses with broken
R-parity is shown as a function of $A_\Delta $ for fixed 
$\sigma_R$.  The soft masses
and right-handed breaking scale,
are of the order of 10 TeV.
The maximum triplet Yukawa coupling allowed by positivity of the mass
eigenvalues in this case is 
$h_\Delta\sim 0.4$.
Even in the maximal case the mass of the
doubly charged scalar $m_{H^{++}}\sim 1 $ TeV.
In Fig. \ref{mpp_ad} b) $m_{H^{++}}$ is plotted  
in the model containing nonrenormalizable terms 
as a function of the nonrenormalizable $b_R$-parameter for 
$v_R^2/M=10^2$ GeV.

\subsection{Doubly charged scalars at linear colliders}

The collider phenomenology of the doubly charged scalars has
been actively studied, since they appear in several extensions of the
Standard Model, can be relatively light and have clear signatures.
The main decay modes for relatively light doubly charged Higgs
are \cite{HMPR}
$H^{--}\rightarrow l_1^-l^-_2$, where $l_{1,2}$ denote leptons.  
Thus the experimental signature of the decay is
a same sign lepton pair with no missing energy, including 
lepton number violating final states.

Since the left-right models contain many extra parameters
when compared to the MSSM, a great advantage of the pair production
is that it is relatively model independent.
The doubly charged Higgses can be produced in 
$f\bar f\rightarrow\gamma^*,Z^*\rightarrow H^{++}H^{--}$ both at
lepton and hadron colliders, if kinematically allowed,
even if $W_R$ is very heavy,  
or the triplet Yukawa couplings are very small.
The pair production cross section at a linear collider has been 
given in \cite{GMS,Gun}.
The cross section remains sufficiently large close to the kinematical
limit for the detection to be possible.

Kinematically,  production of a single doubly charged scalar would be
favoured.
This option is more model dependent, but for reasonable parameter
range the kinematical reach is
approximately doubled compared to the pair production.

\section{Conclusions}

The lightest CP even Higgs boson in SLRM can be considerably heavier 
as compared to the lightest Higgs
in the MSSM, and its couplings to fermions remain similar to the 
couplings of the Standard Model Higgs boson.
In the SLRM with the minimal particle content one has typically also a
light doubly charged Higgs boson.
If this particle is found, it is a strong indication of the SLRM with
minimal particle content.

\section*{References}


\end{document}